\documentclass{IEEEtran}
\usepackage{cite}
\usepackage{amsmath,amssymb,amsfonts}
\usepackage{graphicx}
\usepackage{textcomp,nicefrac}
\usepackage[numbers,sort&compress]{natbib}
\def\BibTeX{{\rm B\kern-.05em{\sc i\kern-.025em b}\kern-.08em
T\kern-.1667em\lower.7ex\hbox{E}\kern-.125emX}}
\begin{document}
\title{Dataset for Neutron and Gamma-Ray Pulse Shape Discrimination}
\author{{Kaimin Wang, Haoran Liu, Peng Li, Mingzhe Liu, Zhuo Zuo}{\thanks{Kaimin Wang is with State Key Laboratory of Geohazard Prevention and Geoenvironment Protection, Chengdu University of Technology, Chengdu, 610059, China. (e-mail: wangKaimin@stu.cdut.edu.cn).}
\thanks{Haoran Liu and Mingzhe Liu are with State Key Laboratory of Geohazard Prevention and Geoenvironment Protection, Chengdu University of Technology, Chengdu, 610059, China, and also with School of Data Science and Artificial Intelligence, Wenzhou University of Technology, Wenzhou 325000, China (e-mail: liuhaoran@cdut.edu.cn; liumz@cdut.edu.cn).}
\thanks{Peng Li is with Engineering \& Technical College of Chengdu University of Technology, Leshan 614000, China (e-mail: lipeng@stu.cdut.edu.cn).}
\thanks{Zhuo Zuo is with Southwest Institute of Physics, Chengdu, 610225, China,and also with Chengdu University of Technology College of Engineering Technology, Leshan, 614000, Sichuan, China (e-mail: zuozhuo@stu.cdut.edu.cn).}}}

\maketitle

\begin{abstract}
This work provides a publicly accessible dataset includes neutron and gamma-ray pulse signals for conducting pulse shape discrimination experiments. Several traditional and recently proposed pulse shape discrimination algorithms are utilized to evaluate the performance of pulse shape discrimination under raw pulse signals and noise-enhanced datasets. These algorithms are zero-crossing (ZC), charge comparison (CC), falling edge percentage slope (FEPS), frequency gradient analysis (FGA), pulse-coupled neural network (PCNN), ladder gradient (LG), and heterogeneous quasi-continuous spiking cortical model (HQC-SCM). In addition to the pulse signals, this dataset includes the source code for all the aforementioned pulse shape discrimination methods. Moreover, the dataset provides the source code for schematic pulse shape discrimination performance evaluation and anti-noise performance evaluation. This feature enables researchers to evaluate the performance of these methods using standard procedures and assess their anti-noise ability under various noise conditions. In conclusion, this dataset offers a comprehensive set of resources for conducting pulse shape discrimination experiments and performance evaluation. It can be accessed at https://doi.org/10.5281/zenodo.7754573.
\end{abstract}


\section{Introduction}
\label{sec:introduction}
\IEEEPARstart{P}{ULSE} shape discrimination (PSD) is a signal processing technique that distinguishes radiation pulse signals based on their shape features. This approach has been applied in various fields, such as surface background rejection \cite{b1}, nuclear and electron recoils discrimination \cite{b2}, and neutron and gamma-ray discrimination \cite{b3}. In particle discrimination applications, different particle injection events are often captured simultaneously by radiation detectors, resulting in an undifferentiated particle count for different types of particles. This phenomenon poses challenges to accurately detecting a specific type of particle. However, the detector response to different particle injection events are not identical. The pulse shapes of different particles exhibit unique features that stem from the intrinsic properties of the interaction between the particles and the detector's sensitive volume. Therefore, PSD exploits these intrinsic pulse shape differences to distinguish particles based on their pulse shape features. The neutron and gamma-ray PSD specifically aims to distinguish between these two types of particle injection events, enabling accurate neutron flux detection. Neutrons are inevitably accompanied by gamma-ray photons due to their interaction with the surrounding environment. As a result, neutron detectors capture both types of particles simultaneously and require PSD to differentiate them. This technique has been extensively utilized in various fields that demand advanced neutron detection and monitoring, including particle and nuclear physics \cite{b4}, meteorology \cite{b5}, astronomy \citep{b6,b7}, and atomic reactors \citep{b8,b9}.

The provided dataset contains radiation pulse signals captured from a neutron and gamma-ray superposed field, as well as source codes for several discrimination methodologies, both traditional and state-of-the-art. This dataset also includes evaluation criteria and the implementation of PSD performance assessment, facilitating easy comparison of the efficacy of different PSD methods. Additionally, anti-noise evaluation codes for PSD methods are included, allowing for validation of each method’s performance across varying noise levels. Overall, this dataset offers a comprehensive source of performance evaluation codes for PSD methodologies. It can be used to compare newly developed PSD methods with existing ones, thereby expediting advancements in the field of PSD.

\section{DATA DESCRIPTION}
\subsection{Neutron and gamma-ray radiation pulse signal}
The folder named ‘Neutron and Gamma Radiation Pulse Signal’ provides three types of signals: raw, filtered, and noise-enhanced signals, each containing neutron and gamma-ray pulse signals. The raw signal was obtained using a \textsuperscript{241}\textrm{Am}-Be isotope neutron source that generated a superposed field of neutrons and gamma-ray photons. To collect these pulse signals, an EJ299-33 plastic scintillator and a digital oscilloscope (TPS2000B) with a 1 GS/s sampling rate, 8 bits vertical resolution, and 200 MHz bandwidth were used. The trigger threshold was set at 500 mV to ensure reliable data collection, corresponding to an energy of approximately 1.6 MeV as defined in \cite{b10}. To avoid information corruption within the signals, a pulse duration of 280 ns was carefully selected based on the Shannon criteria \cite{b11}. Typical neutron and gamma-ray pulse signals are shown in Fig. 1. 
\begin{figure}[t]
\centerline{\includegraphics[width=3.5in]{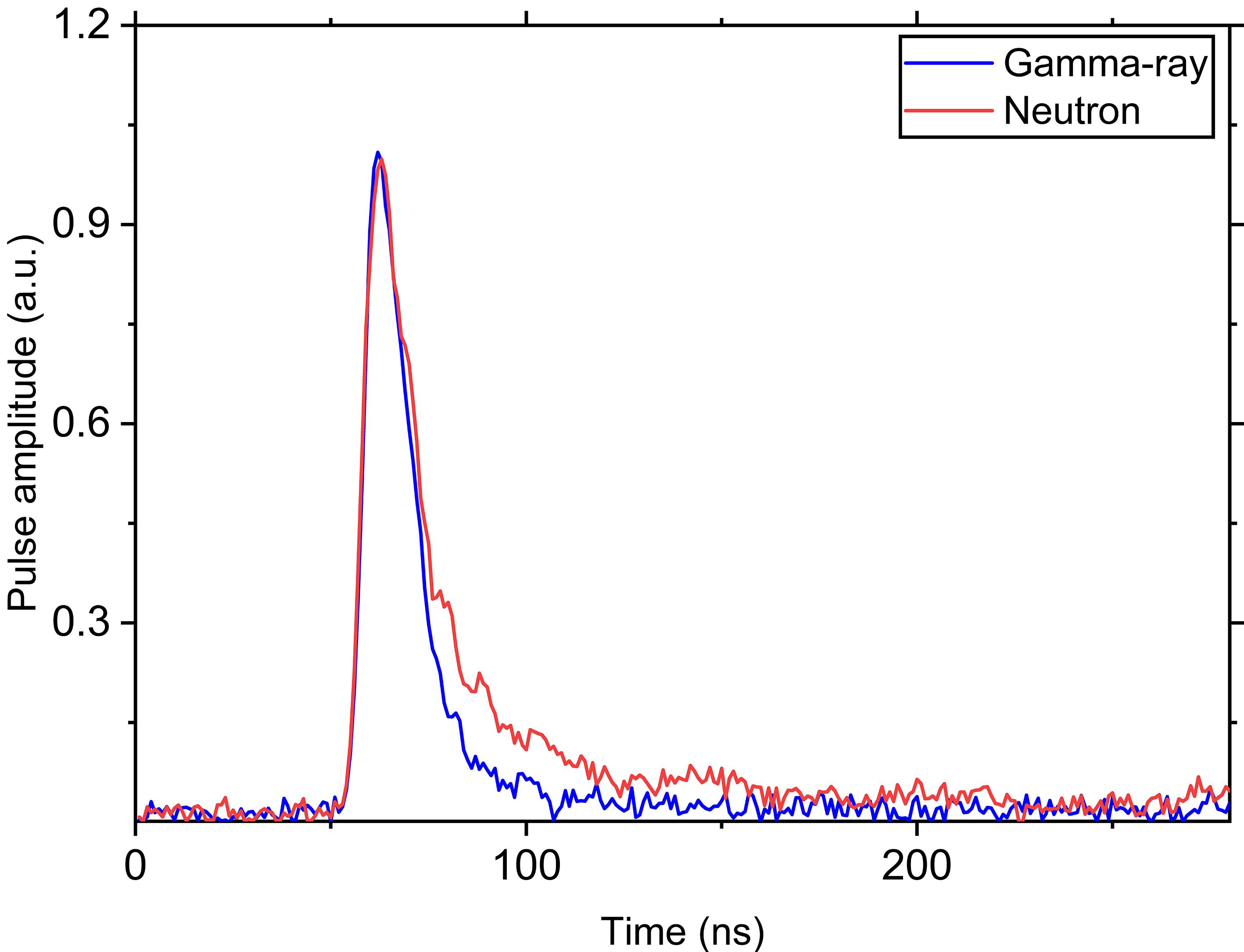}}
\caption{\centering Typical neutron and gamma-ray pulse signals.}
\label{fig1}
\end{figure}
The differences between these two types of particles’ pulses locates at the falling edge and delayed fluorescence parts, ranging from approximately 80-100 ns and 100-150 ns, respectively.The filtered signal was obtained by applying a Fourier filter to remove low-frequency noise from the raw signal, which is a standard pre-processing step for PSD applications. This step improves the performance of discrimination methods by making them work in low-noise scenarios. Finally, the noise-enhanced signal was obtained by adding Gaussian noise with a variance of 0.001 to the raw signal, simulating extreme noise conditions. This file was provided to help evaluate the performance of PSD methods in high-noise scenarios.
\subsection{Discrimination methods}
The folder named “Discrimination methods” contains the source codes for seven different methodologies used for neutron and gamma-ray PSD. These PSD methods include zero-crossing (ZC), charge comparison (CC), falling edge percentage slope (FEPS), frequency gradient analysis (FGA), pulse-coupled neural network (PCNN), ladder gradient (LG), and heterogeneous quasi-continuous spiking cortical model (HQC-SCM).

To perform PSD on the raw signal, simply run the “main.m” program using the default settings, which utilizes all PSD methods. The discrimination factors of each method are then calculated and used to generate histograms. An automatic double Gaussian distribution fitting process is then applied, where the neutron distribution is located on the right side of the histogram and the gamma-ray distribution is located on the left side. The three-sigma points of each Gaussian distribution are used as the end of the distribution since they contain 99.74\% of the distribution. The central point between the right side three-sigma point of the gamma-ray distribution and the left side three-sigma point of the neutron distribution is used as the dividing point between gamma-rays and neutrons. It should be noted that all PSD methods, except HQC-SCM and ZC, require manual parameter settings. By default, the parameters are optimized to achieve near-optimal performance for the given dataset. However, when using other datasets, these parameters must be adjusted accordingly. Conversely, ZC is a parameter-free method that requires no additional tuning. Additionally, HQC-SCM incorporates a genetic algorithm-based automatic parameter selection approach, which allows for direct implementation on other datasets.

The “main.m” program outputs the discrimination results of all PSD methods based on the histograms and dividing points, with each pulse signal categorized as either a 0 for gamma-rays or a 1 for neutrons. The results are saved in a mat file called ‘Output.mat’.

These discrimination algorithms can be easily implemented on other neutron and gamma-ray pulse signals by changing the ‘raw\_signal.txt’ file. This includes signals with enhanced noise, filtered de-noised signals, and signals from other radiation detection systems. 
\subsection{Discrimination performance evaluation}
The folder named 'Discrimination performance evaluation' includes the source codes for comparing the PSD performance of the seven methodologies mentioned earlier. To evaluate the discrimination performance, one can run the 'main.m' program with the default settings. This program performs discrimination factor calculation, histogram generation, and double Gaussian fitting for all discrimination methods discussed in the previous section. After this, the figure of merit (FOM) value is calculated for each discrimination method using the Gaussian fitting results and the formula given below,
\begin{equation}
\label{eq1}
FOM = \dfrac{S}{FWHM_{n}+FWHM_{\gamma}},
\end{equation}

where, $ FWHM_{n} $ and$ FWHM_{\gamma} $ represent the full width at half maximum of the neutron Gaussian group and gamma-ray Gaussian group, respectively. A large distance S and small values of$ FWHM_{n} $ and$ FWHM_{\gamma} $ indicate good discrimination performance. Consequently, a higher FOM value indicates a better discrimination performance.

The ‘main.m’ program outputs the histogram of each PSD method’s discrimination factor, together with the FOM value calculated from this histogram.

As shown in Fig. 2, each PSD method's histogram exhibits two Gaussian distributions. The left side distribution contains the counts of gamma-ray pulse signals, and the other one contains the counts of neutron pulse signals.

A large distance between these two groups, and a small Gaussian distribution variance help separate neutron and gamma-ray pulse counts. The histogram and FOM value give subjective and objective PSD performance evaluations.
\begin{figure*}[t]
	\centering
	\includegraphics[width=\textwidth]{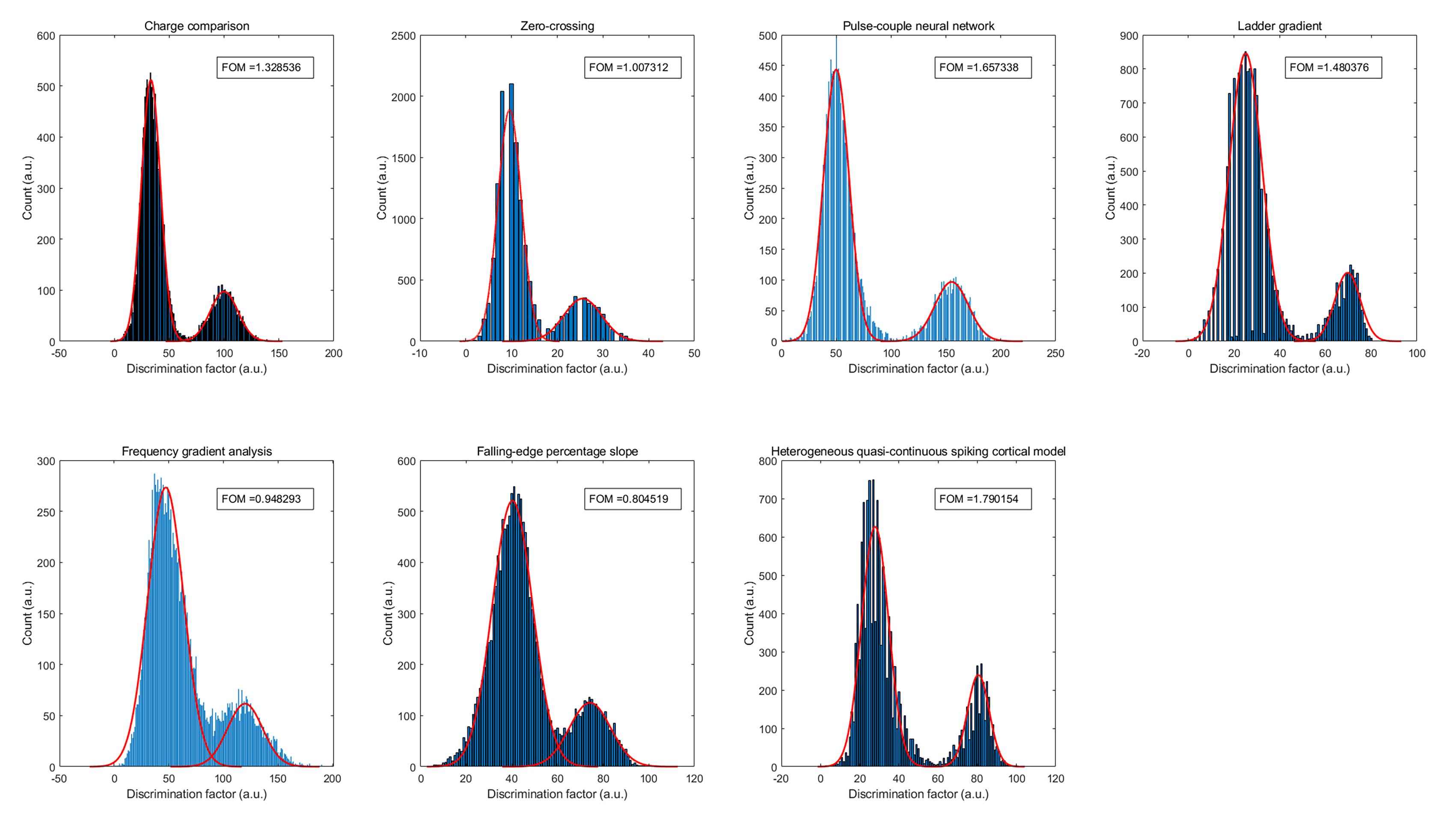}
	\caption{Discrimination performance evaluation}
\label{fig2}
	\centering
	\includegraphics[width=\textwidth]{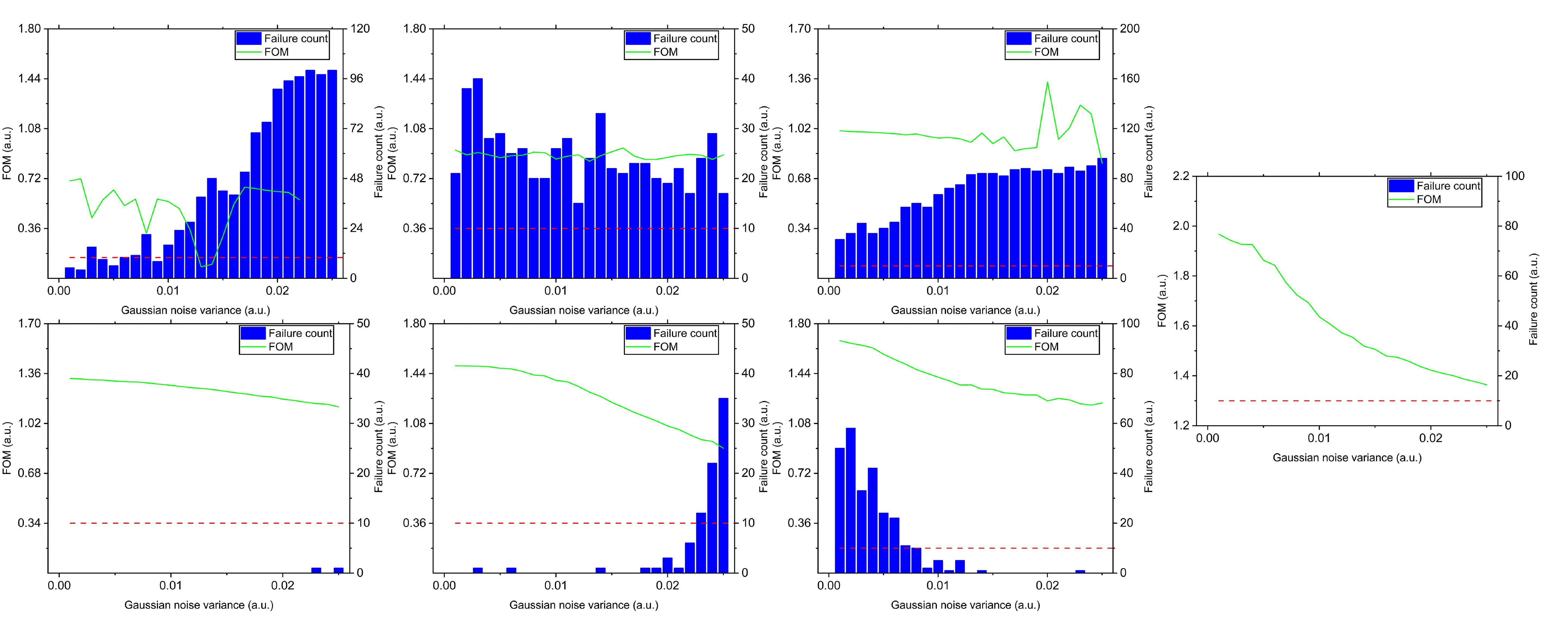}
	\caption{Anti-noise performance evaluation}
\label{fig3}
\end{figure*}
\subsection{Anti-noise experiment}
The folder named ‘Anti-noise experiment’ contains the source code of PSD performance evaluation of all seven PSD methods previously mentioned in various noise scenarios.

To evaluate the anti-noise ability of the PSD methodologies, one can simply run the ‘main.m’ program, which performs PSD experiments on datasets consisting of raw neutron and gamma-ray pulse signals with additional Gaussian noise. The variance of the added noise ranges from 0.01 to 0.025. Discrimination processes are independently executed one hundred times for each method and noise level to obtain the average performance of each method under different noise conditions.

In high noise scenarios, many PSD methods fail to generate histograms that can be fitted by Gaussian distribution. Hence, the failure count of discrimination methods is recorded. If a method's failure count exceeds ten under a noise level, it is deemed unreliable under that noise condition.

The "main.m" program outputs two mat files that contain FOM values, averaged FOM values, and failure counts of all the discrimination methods mentioned before. Based on this information, the performance of each discrimination method can be visualized as shown in Fig. 3.
\begin{figure}[t]
	\centering
	\includegraphics[width=3.5in]{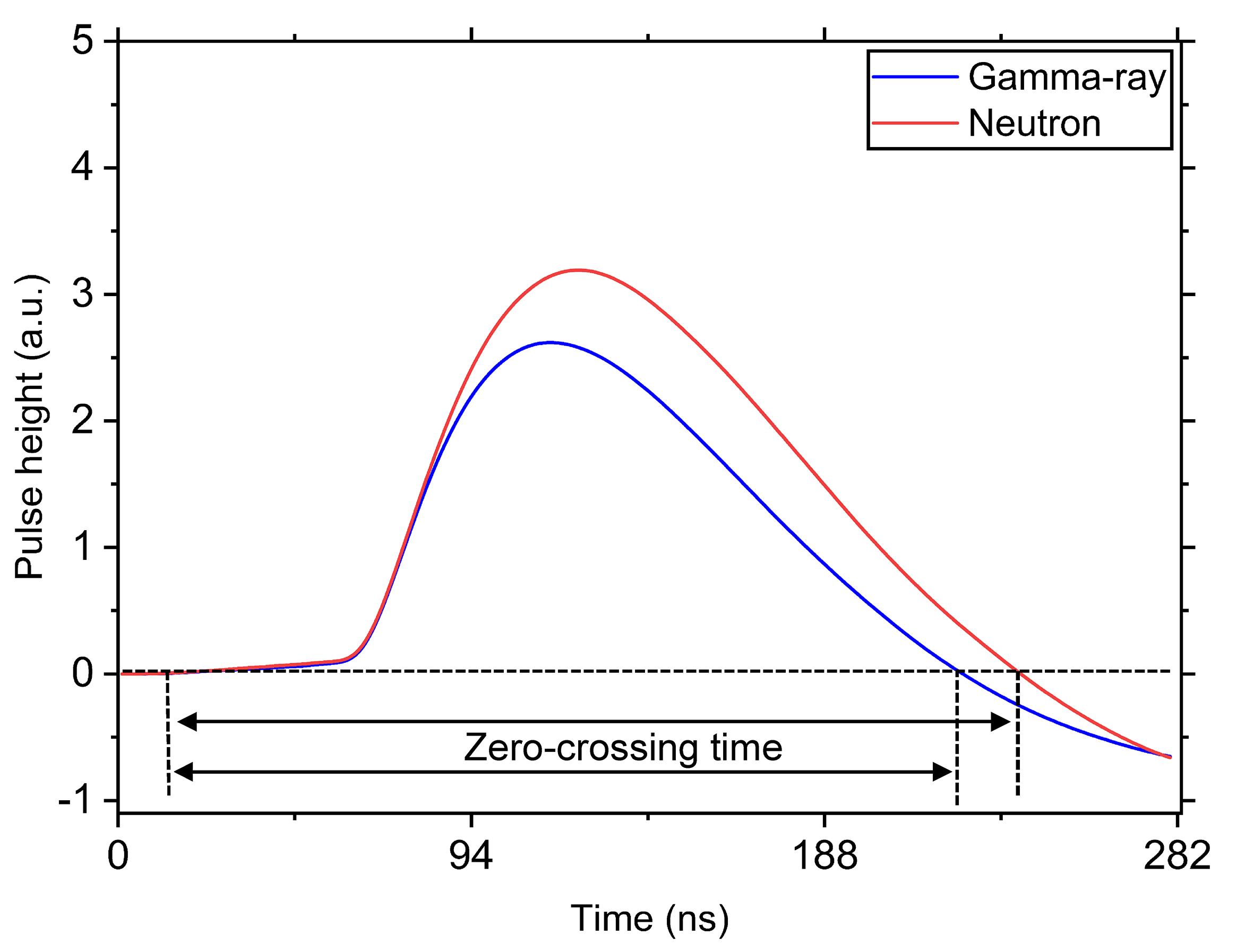}
	\caption{Schematic diagram of the zero-crossing}
\label{fig4}
\end{figure}
\begin{figure}[t]
	\centering
	\includegraphics[width=3.5in]{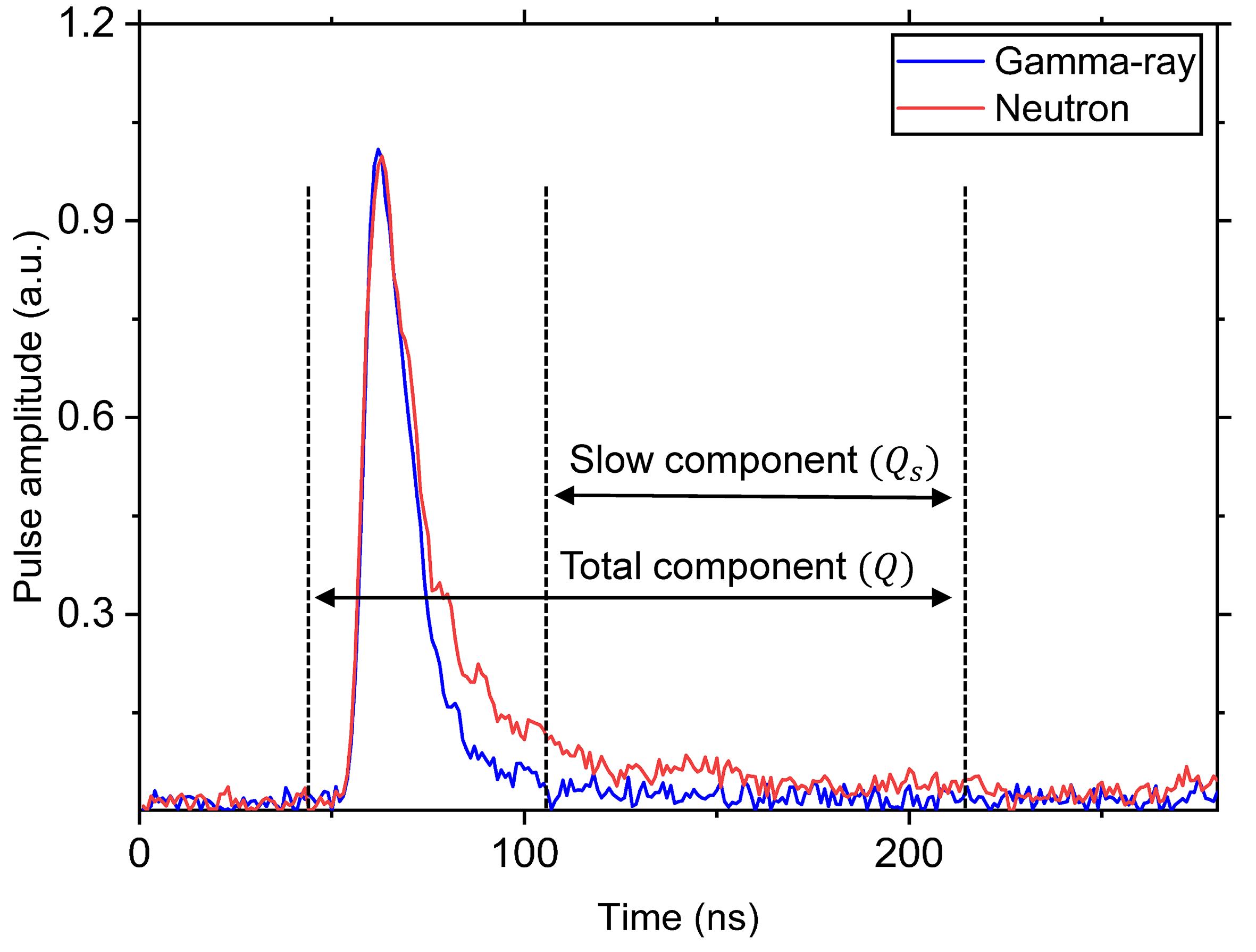}
	\caption{Schematic diagram of the charge comparison}
\label{fig5}
\end{figure}
Fig. 3 displays that the FOM values of all PSD methods decrease as the noise level increases, and the failure counts of most PSD methods tend to increase as the noise becomes more intense. However, recently developed PSD methodologies like PCNN and HQC-SCM outperform traditional methods such as ZC and FGA with higher average FOM values and fewer failure counts.

\section{METHODOLOGIES}
\subsection{Zero-crossing}
The Zero-crossing (ZC) method is a widely used recursive algorithm in digital signal processing \cite{b12} , particularly in particle PSD due to its simple implementation and low computational complexity. This method involves passing the pulse signal through a virtual differential-integral-integrator (CR-RC2) network to produce a bipolar pulse signal, as shown in Fig. 4 Neutrons decay at a slower rate in scintillators than gamma-rays, causing a longer zero-crossing time $R_{ZC}$for neutrons compared to gamma-rays. This difference is utilized to distinguish between neutrons and gamma-rays. The mathematical expression of CR-RC2 is defined as follows,
\begin{align}
y[n] &= 3\delta^2 y[n-1]-3\delta y[n-2]+\delta^3 y[n-3]\nonumber\\
 &+T\delta (1-\dfrac{\omega T}{2})x[n-1]-T\delta^2 (1+\dfrac{\omega T}{2})x[n-2],
\end{align}

where, $y$ represents the bipolar pulse signal obtained through CR-RC2; $x$ represents an original neutron and gamma-ray pulse signal; $n$ is the sampling index of the signal; and $\delta$ and $\omega$ are two constants defined as follows,
\begin{equation}
\label{eq3}
\delta=e^{-\frac{T}{\tau}},
\end{equation}
\begin{equation}
\label{eq4}
\omega=\frac{1}{\tau},
\end{equation}

where, the sampling time interval is represented by $T$; and the shaping time of the signal is represented by $\tau$=RC.
\subsection{Charge comparison}
The charge comparison (CC) method \cite{b13} is currently the most commonly used algorithm for PSD, owing to its stability, reliability, and simplicity in calculation. The fluorescence pulses generated by the scintillator exhibit distinctive variations for different incident particles due to their diverse interactions. As depicted in Fig. 5 the segment of the pulse signal produced
by the slow luminescence of the scintillator is termed the slow component, while the entire pulse signal is referred to as the total component. The dis-crimination between neutrons and gamma-rays is based on the difference in the charge ratio between the slow and total components of their fluorescence pulses. The mathematical expression for this ratio is defined as follows,
\begin{equation}
\label{eq5}
R_{CC} = \frac{Q_{s}}{Q},
\end{equation}

where, the slow component charge amount and total component charge amount are represented by $Q_{s}$ and $Q$, respectively. Neutrons decay more slowly than gamma-rays and exhibit a unique delayed fluorescence effect, resulting in a higher ratio $R_{CC}$ for neutrons compared to gamma-rays.
\subsection{Falling edge percentage slope}
The falling edge percentage slope (FEPS) method is designed for rapid discrimination of neutrons and gamma-rays \cite{b14}. To apply this method, a region of interest (ROI) must be selected within the pulse signal region where the differences between neutrons and gamma-rays are most pronounced, namely the falling edge region. As illustrated in Fig. 6 the ROI is identified by defining upper threshold (UT) and lower threshold (LT) values, and the intersection points between the pulse signal and the upper and lower threshold values are determined to obtain coordinates ($\psi_{x}$,$\psi_{y}$)and($\phi_{x}$,$\phi_{y}$). The discrimination factor is defined as follows,
\begin{equation}
\label{eq6}
R_{FEPS} = \frac{\psi_{y}-\phi_{x}}{\psi_{x}-\phi_{y}},
\end{equation}

Furthermore, the LT is fixed at 10\% of the peak value of the pulse signal, whereas the UT is a parameter that requires manual configuration and can range between 30\% and 90\% of the pulse signal's peak value. As the light pulse generated by the interaction of gamma-rays with scintillators has a shorter decay time than that of neutrons, the absolute value of $R_{FEPS}$ for gamma-rays will surpass that for neutrons.
\begin{figure}[t]
	\centering
	\includegraphics[width=3.5in]{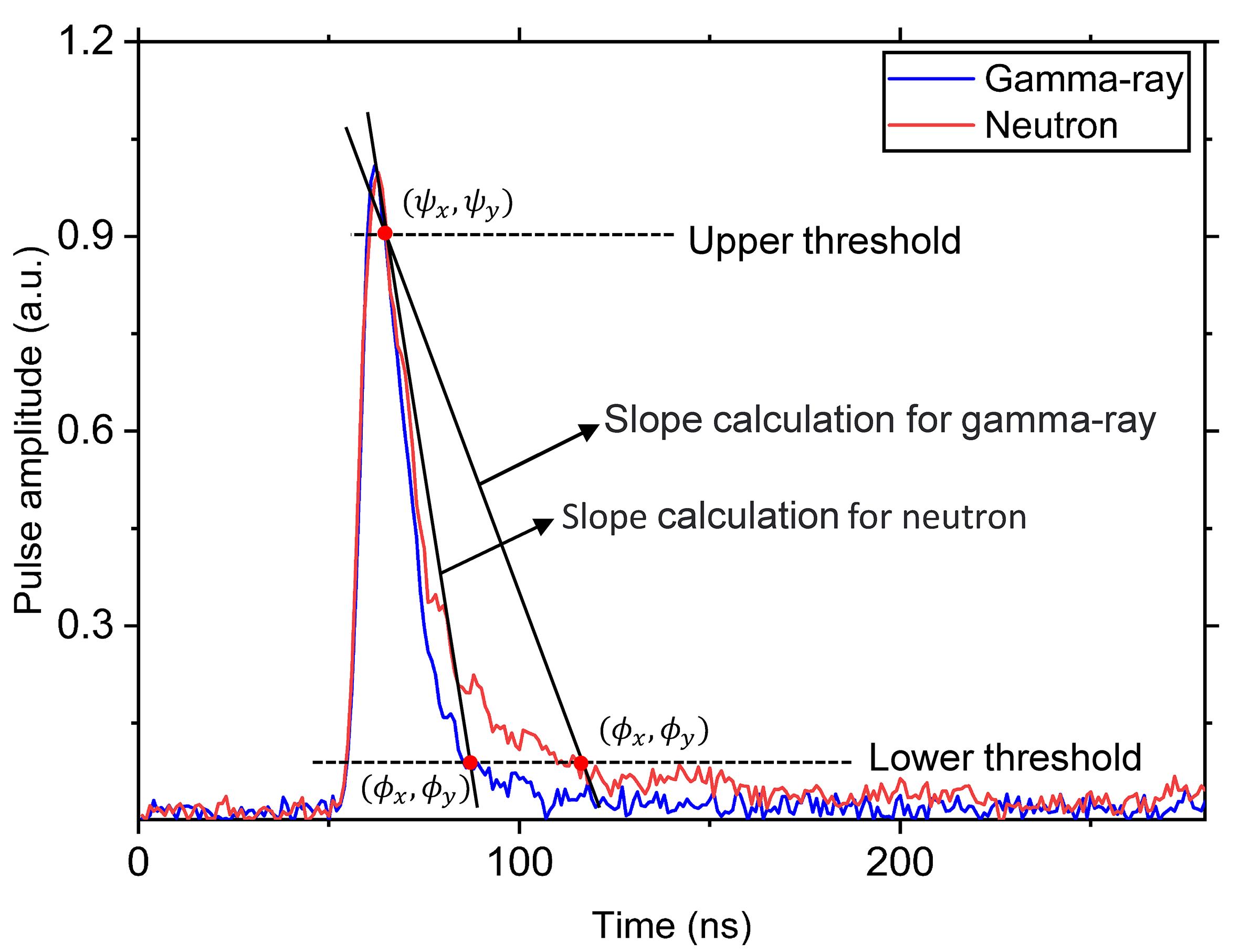}
	\caption{Schematic diagram of the falling edge percentage slope}
\label{fig6}
\end{figure}
\begin{figure}[t]
	\centering
	\includegraphics[width=3.5in]{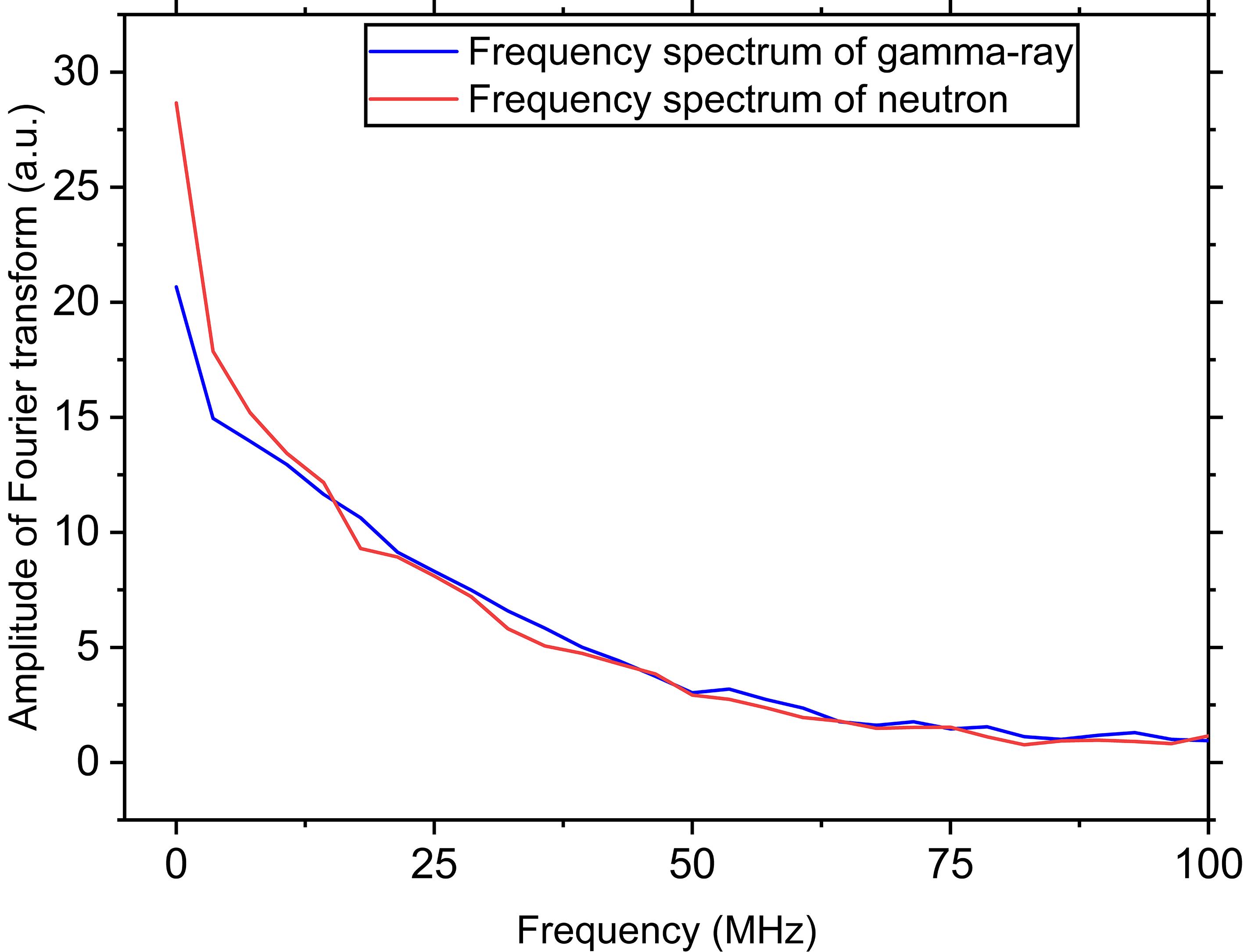}
	\caption{Schematic diagram of the frequency gradient analysis}
\label{fig7}
\end{figure}
\begin{figure}[t]
	\centering
	\includegraphics[width=3.5in]{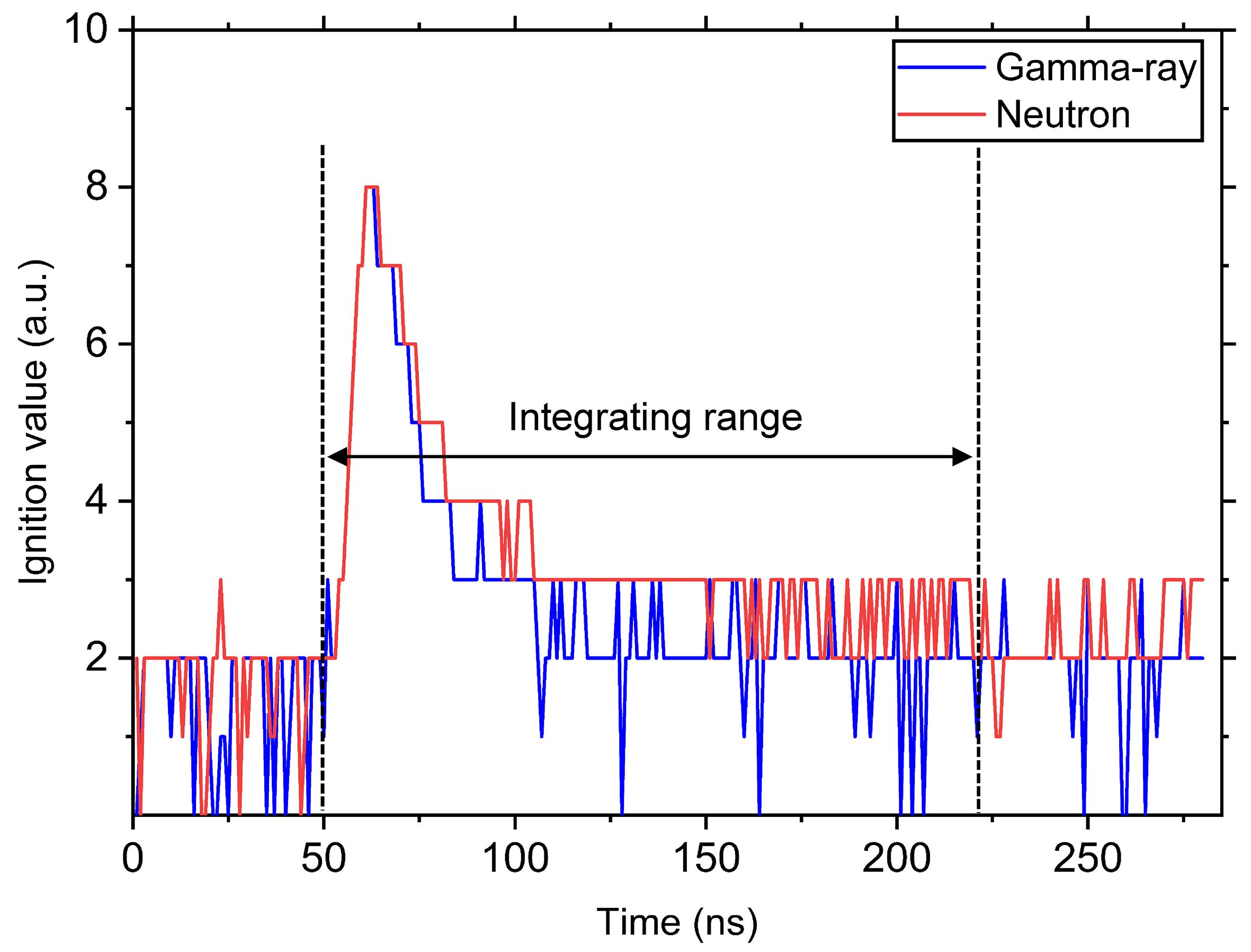}
	\caption{Schematic diagram of the pulse-coupled neural network}
\label{fig8}
\end{figure}
\subsection{Frequency gradient analysis}
The frequency gradient analysis (FGA) method is a technique proposed by Liu et al. in 2010 \cite{b15} for discrimination of neutrons and gamma-rays based on frequency analysis. This method is known for its ability to resist noise. The spectra of neutron and gamma-ray pulses generated by the fast Fourier transform are shown in Fig. 7 It is evident from the figure that the two types of pulses exhibit a significant difference at zero frequency, while the difference is less pronounced at higher frequencies. FGA exploits this characteristic to distinguish between neutrons and gamma-rays. The mathematical formula for this method is given by,
\begin{equation}
\label{eq7}
R_{FGA}= \frac{\left|{\left| X(0) \right|-\left| X(f) \right|}\right|}{f},
\end{equation}

where, $\left| X(0) \right|$ and $\left| X(f) \right|$ are the amplitude values of the Fourier transforms at frequencies of 0 and $f$, respectively. Since the amplitude of neutrons at frequency 0 in the Fourier domain is greater than that of gamma-rays, the $R_{FGA}$ value for neutrons is higher than that for gamma-rays.
\subsection{Pulse-coupled neural network}
The pulse-coupled neural network (PCNN) method is a novel method for pulse shape discrimination, developed by Liu et al. in 2021 \cite{b16}. PCNN is capable of capturing and recognizing dynamic information in the neutron and gamma-ray pulse signals, which is critical in the discrimination process. The method is particularly efficient in dealing with noise and significantly outperforms traditional discrimination methods. The unique structure and working mechanism of PCNN are inspired by the synchronized pulse firing and global coupling observed in the animal visual cortex.

The cortical cells in the brain produce and transmit stimulating signals between cell assemblies when an animal's eyes are stimulated by external light, conveying the image captured by the eyes\citep{b17,b18}. Through natural selection, this mechanism has been highly effective in processing dynamic information in images \cite{b19}.
\begin{figure}[t]
	\centering
	\includegraphics[width=3.5in]{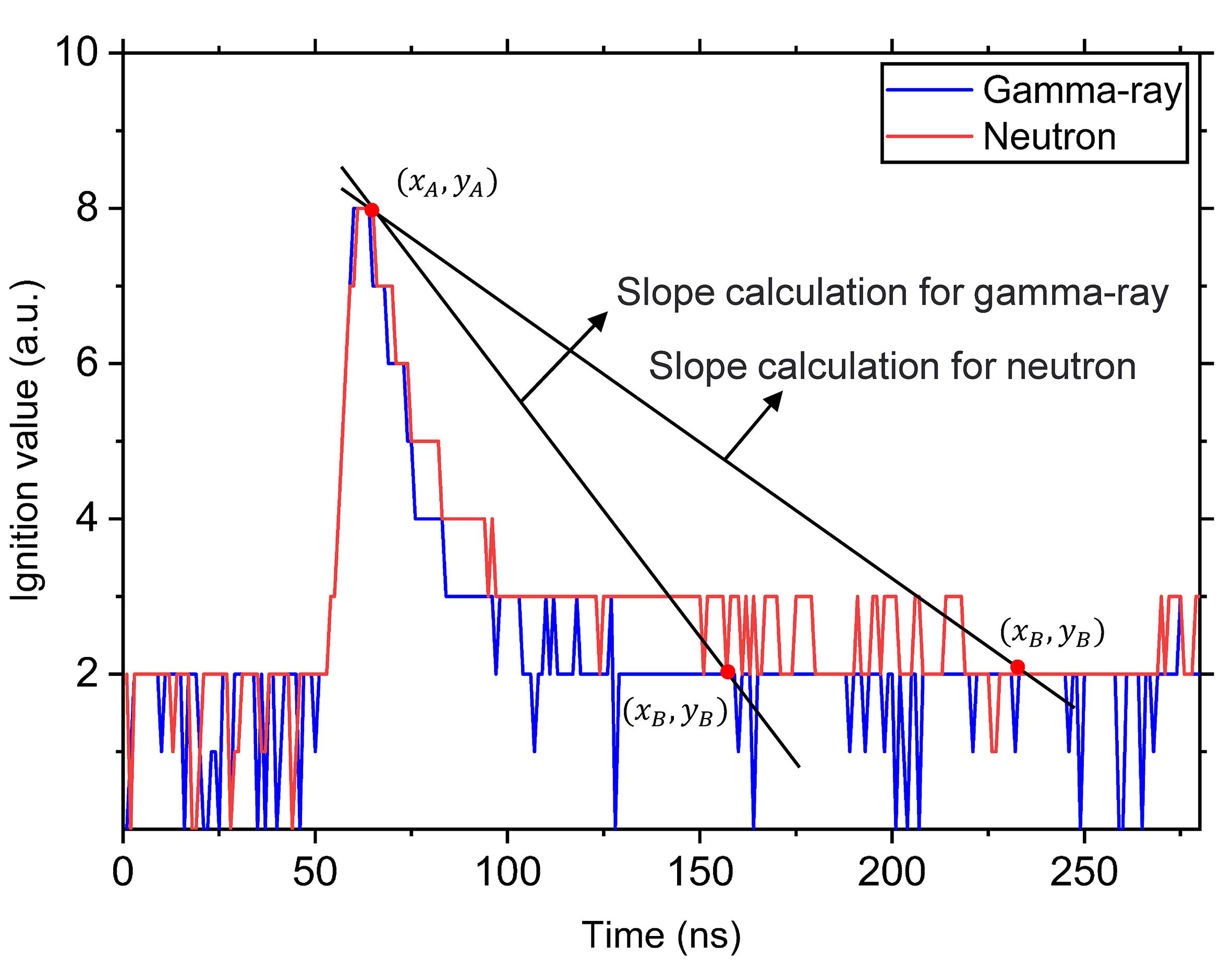}
	\caption{Schematic diagram of the ladder gradient}
\label{fig9}
\end{figure}
\begin{figure}[t]
	\centering
	\includegraphics[width=3.5in]{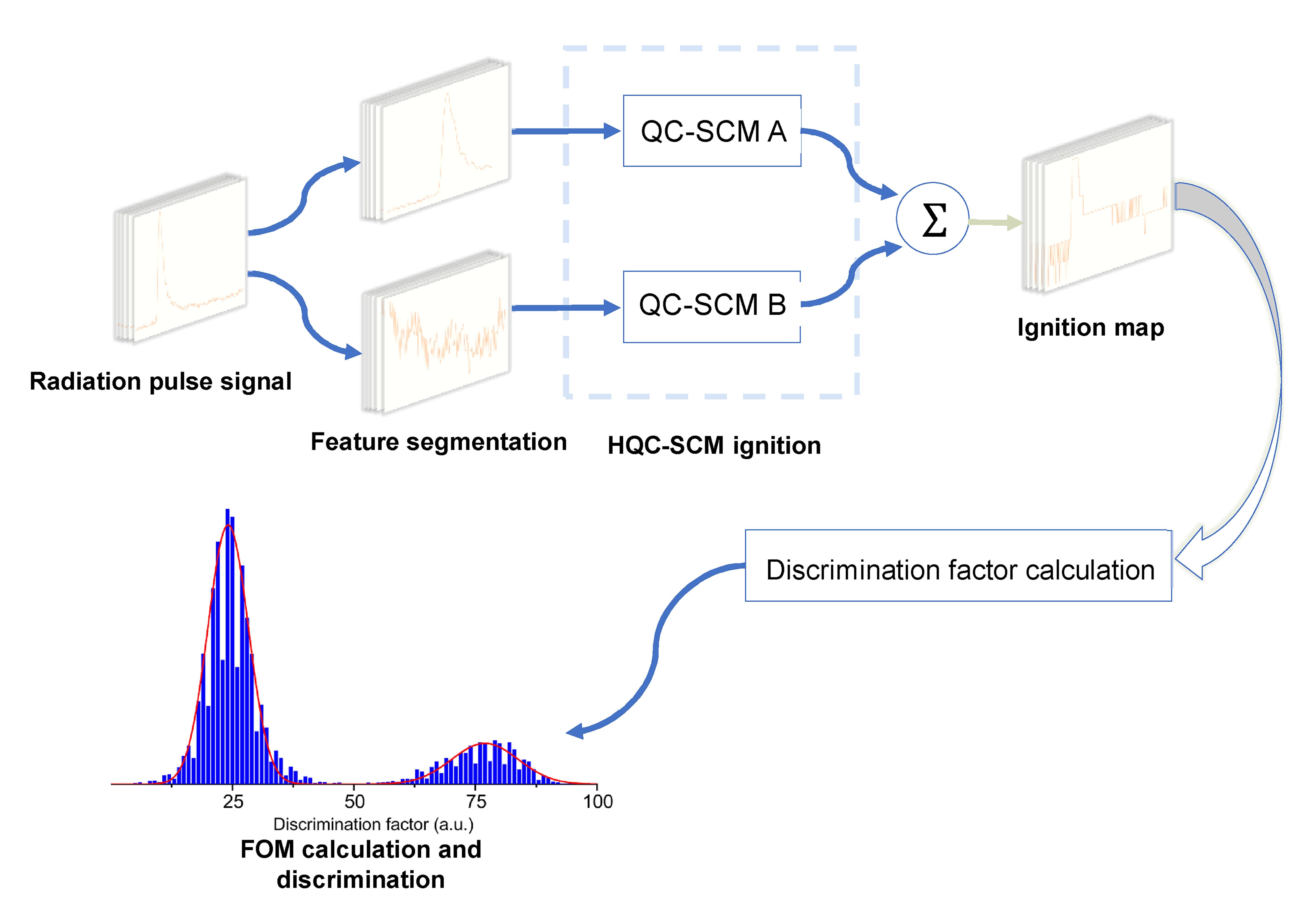}
	\caption{Schematic diagram of the heterogeneous quasi-continuous spiking cortical model}
\label{fig10}
\end{figure}
Based on these neurobiological findings, researchers designed PCNN specifically for image processing \citep{b20,b21}, which has been widely used in image processing fields such as segmentation \cite{b20}, pattern recognition \cite{b22}, and feature extraction \cite{b21}.The PCNN structure comprises three components: the receptive field, the modulated field, and the pulse firing field. The receptive field is composed of linked input (LI) and feedback input (FI) modules, with LI receiving stimulation from surrounding neurons and FI being primarily influenced by external stimuli. The pulse firing field couples FI and LI and emits pulses based on the internal activity of neurons. These three components interact and constrain each other to determine the system's output. The internal activity of PCNN can be expressed mathematically using the following formula,
\begin{equation}
\label{eq8}
	F_{ij}[n] = V_{F} \sum_{kl} M_{ijkl} Y_{kl}[n-1] + S_{ij}\\
	+e^{-\alpha_{F}}F_{ij}[n-1],
\end{equation}
\begin{equation}
\label{eq9}
    L_{ij}[n] = e^{-\alpha_{L}} L_{ij}[n-1]+V_{L}\sum_{kl}W_{ijkl}Y_{kl}[n-1],
\end{equation}
\begin{equation}
\label{eq10}
	U_{ij}[n] = F_{ij}[n] (1 + \beta L_{ij}[n]),
\end{equation}
\begin{equation}
\label{eq11}
	\theta_{ij} [n] = e^{-\alpha_{\theta}}\theta_{ij}[n-1]+V_{\theta}Y_{ij}[n-1],
\end{equation}
\begin{equation}
\label{eq12}
	Y_{ij}[n] = \left\{
	\begin{array}{l}
		1, U_{ij}[n] > \theta_{ij}[n], \\
		0, \text{ otherwise}
	\end{array}
	\right.
\end{equation}

where,$F_{ij}$ represents the activity level of FI for the neurons located at coordinates i and j; n is the iteration number; $S_{ij}$ denotes external stimulation; $L_{ij}$ represents the activity level of LI; $V_{F}$ and $V_{L}$ denote the amplification factors of FI and LI; $W_{ijkl}$ and $M_{ijkl}$ denote the coupling weight matrixes between neurons and their surrounding neurons; $U_{ij}$ is the internal activity threshold, modulated by both FI and LI; $\theta_{ij}$ is the dynamic threshold; $V_{\theta}$ represents the amplification factor of the dynamic threshold; $\beta$ represents the linking strength, which primarily regulates the impact of $L_{ij}$ on $U_{ij}$; and $Y_{ij}$ represents the pulse fired by the neuron, depending on the relationship between $U_{ij}$ and $\theta_{ij}$. When $U_{ij}$ exceeds $\theta_{ij}$, the neuron is activated, denoted by $Y_{ij}$=1. Subsequently, $\theta_{ij}$ is amplified $V_{\theta}$-fold to suppress the neuron's activation in the next iteration. Furthermore, $\alpha_{F}$, $\alpha_{L}$, and $\alpha_{\theta}$ denote the attenuation constants of FI, LI, and $\theta_{ij}$, respectively. These parameters simulate the characteristics of the biological neuron membrane potential decay, enabling the neuron to return to the resting level after being activated.

In order to use the PCNN model for neutron and gamma-ray discrimination, the pulse signals must be input into the PCNN to generate ignition maps, with each pulse signal corresponding to a distinct ignition map. As depicted in Fig. 8 the differences between neutron and gamma-ray pulses in the falling edge and delayed fluorescence parts are magnified in the ignition maps, making it easier to distinguish between the two particle types. By integrating the ignition map from approximately 15 ns 
before to 125 ns after the pulse peak, which contains the falling edge and delayed fluorescence portions, the discrimination factor $R_{PCNN}$ is obtained. The $R_{PCNN}$ value for neutrons is notably larger than that for gamma-rays.
\subsection{Ladder gradient}
The ladder gradient (LG) method is based on the quasi-continuous spiking cortical model (QC-SCM) \cite{b23}. This method offers several advantages over previous techniques, including the ability to more accurately identify and capture dynamic information in neutron and gamma-ray signals, as well as increased resistance to noise. These advantages stem from LG's use of a non-integer iteration step in QC-SCM, resulting in higher resolution. However, these advantages come at the cost of increased computational complexity.

To address the issue of computational complexity, LG employs a direct gradient calculation method to compute the discrimination factor $R_{LG}$, rather than the integration method used in CC and PCNN. This approach significantly reduces the computational complexity of discrimination, making LG more suitable for a wider range of discrimination tasks. The mathematical expression of QC-SCM is as follows,
\begin{equation}
\label{eq13}
	U_{ij}(t+\Delta t) = f^{\Delta t}U_{ij}(t)+S_{ij}(1+\sum_{kl}W_{ijkl}Y_{kl}(t))
\end{equation}
\begin{equation}
\label{eq14}
   \theta_{ij}(t+\Delta t) = g^{\Delta t}\theta_{ij}(t)+hY_{ij}(t)
\end{equation}
\begin{equation}
	Y_{ij}(t+\Delta t) =
	\begin{cases}
	\label{eq15}
	1,  & \text{if}\ \frac{1}{1+exp[-({U_{ij}}(t+\Delta t)-{\theta_{ij}}(t+{\Delta t}))]}>\frac{1}{2} \\
	0, & \text{otherwise}
	\end{cases}
\end{equation}
\begin{equation}
\label{eq16}
	R_{LG} = \frac{y_{A}-y_{B}}{x_{A}-x_{B}}
\end{equation}

where, $U_{ij}$ represents the internal activity threshold, where $i$ and $j$ denote the position coordinates of the neuron; $t$ represents the iteration number; $\Delta t$ is a feature parameter of QC-SCM, taking values in the range of 0-1. the closer $\Delta t$ is to 0, the closer the system is to a continuous-time system, hence has better signal processing resolution; $S_{ij}$ represents the external stimulus; $W$ is the weight matrix of the neuron with surrounding neurons, indicating the contribution value of the surrounding neurons; $\theta_{ij}$ represents the dynamic threshold; $h$
is the amplification factor of $\theta_{ij}$;$Y$ represents the firing of the neuron, with $Y$=1 when $U_{ij}$ is greater than  $\theta_{ij}$; $\theta_{ij}$ is amplified by $h$ to inhibit the next firing of the neuron; and $f$ and $g$ are the decay constants of $U_{ij}$ and  $\theta_{ij}$, respectively, used to simulate the process of a biological neuron returning to its resting level after being activated.

The LG method's discrimination principle is illustrated in Fig. 9 The pulse signal is input into the QC-SCM to generate an ignition map. By comparing the LG's ignition map with that of the PCNN (Fig. 8, it is clear that the LG exhibits stronger resistance to noise in the pulse signal while capturing the dynamic information in the signal. The LG calculates the discrimination factor $R_{LG}$ using a ladder gradient calculation approach. Firstly, the positions of the highest point and the m-th mode after the highest point in the ignition map are selected as  ($x_{A}$,$y_{A}$) and ($x_{B}$,$y_{B}$) respectively. Secondly, the slope between these two points is used as the discrimination factor, as shown in formula (\ref{eq15}). It is evident that the discrimination factor of neutrons is greater than that of gamma-rays.
\subsection{Heterogeneous quasi-continuous spiking cortical model}
The heterogeneous quasi-continuous spiking cortical model (HQC-SCM) is a novel algorithm for discriminating neutron and gamma-ray pulse shapes that builds on the previous third-generation neural network-based methods. Unlike the PCNN and LG methods, which require researchers to manually select feature parameters, the HQC-SCM algorithm automatically selects the optimal parameters using a genetic algorithm (GA). This approach eliminates the need for researchers to rely on their experience to choose the appropriate feature parameters, which can lead to insufficient feature extraction and poor discrimination performance.

The HQC-SCM algorithm also addresses the issue of homogeneous neural structures in PCNN and QC-SCM, which may not fully utilize the information contained in pulse signals with multiple types of features. To address this issue, HQC-SCM employs two individual models to extract the features of the falling edge and delayed fluorescence parts of the pulse signal. This method enables the HQC-SCM algorithm to fully extract the features of neutron and gamma-ray pulse signals and achieve outstanding discrimination performance, particularly in noisy environments.

Fig. 10 illustrates the signal processing steps of the HQC-SCM algorithm for pulse shape discrimination. The incoming pulse signal is initially divided into two parts, containing the falling edge and delayed fluorescence features, respectively, using the 5\% position of the peak value of the average pulse signal as the separation point. The falling edge part is processed by QC-SCM A, while the delayed fluorescence part is handled by QC-SCM B. The ignition maps obtained from both models are then merged, and the discrimination factor $R_{HQC-SCM}$ is calculated based on the integrated feature-rich regions, similar to $R_{PCNN}$. The $R_{HQC-SCM}$ value for neutrons is higher than that of gamma-rays. Experimental results have shown that HQC-SCM effectively extracts the characteristic information of pulse signals while demonstrating robustness against noise. QC-SCM A efficiently extracts and enhances the falling edge information, while QC-SCM B effectively captures the information in the delayed fluorescence region. The merged ignition map shows significantly lower high-frequency fluctuations in the delayed fluorescence effect region compared to LG. HQC-SCM outperforms LG and PCNN in pulse shape discrimination.
\section*{Appendix}
The dataset associated with this article is openly available at Zenodo with the following DOI: https://doi.org/10.5281/zenodo.7754573. MATLAB R2022B is the recommended test environment for this dataset. This work is licensed under a Creative Commons Attribution 4.0 License.

\end{document}